\documentstyle[prl,aps,epsf,twocolumn]{revtex}

\begin{document}

\draft

\twocolumn[\hsize\textwidth\columnwidth\hsize\csname@twocolumnfalse%
\endcsname


\title{Hidden Order in the Cuprates}

\author{Sudip Chakravarty$^1$, R. B. Laughlin$^2$, Dirk K. Morr$^3$,
        Chetan Nayak$^1$}

\address{$^1$ Physics Department, University of California Los Angeles,
         Los Angeles, CA 90095--1547\\
         $^2$Department of Physics, Stanford University,
         Stanford, CA 94305\\
         $^3$ Theoretical Division, Los Alamos National Laboratory,
         Los Alamos, NM 87545}

\date{\today}

\maketitle

\begin{abstract}

We propose that the enigmatic pseudogap phase of cuprate superconductors
is characterized by a hidden broken symmetry of $d_{x^2-y^2}$-type.  The
transition to this state is rounded by disorder, but in the limit that the
disorder is made sufficiently small, the pseudogap crossover should reveal
itself to be such a transition.  The ordered state breaks time-reversal,
translational, and rotational symmetries, but it is invariant under the
combination of any two.  We discuss these ideas in the context of ten
specific experimental properties of the cuprates, and make several
predictions, including the existence of an as-yet undetected metal-metal
transition under the superconducting dome.
\end{abstract}

\vspace{1 cm}

\vskip -0.5 truein


]
\narrowtext

\section{Introduction}

In this paper we argue that much of the strange phenomenology of the
cuprate superconductors may be simply explained as the disorder-frustrated
development of a new order parameter.  There are a number of potential
candidates for this order, but the one we favor on phenomenological
grounds is orbital antiferromagnetism \cite{halperin} or $d$-density wave
(DDW) order \cite{schulz,nayak}, which is characterized by a local order
parameter which distills the universal physics underlying the staggered
flux state \cite{hsu,affleck,kotliar,palee,wen} divorced from the
uncontrolled approximations associated with the gauge theory formalism.
The essence of our idea is that the pseudogap\cite{timusk} observed in
underdoped cuprates is an actual gap in the one-particle excitation
spectrum at the wavevector $(\pi, 0)$ and symmetry-related points of the
Brillouin zone associated with the development of this new order.  It is
``pseudo'' in experiment only because of extreme sensitivity to sample
imperfection caused by proximity to the phase transition. Moreover, the
DDW couples weakly to common experimental probes, and is thus difficult to
detect.

Our proposal has much in common with theoretical ideas already in the
literature
\cite{wen,timusk,varma,emery,kivelson,caprara,schmalian,anderson,zhang,balents,randeria,chen},
and borrows heavily from them. For example, Wen and Lee have proposed
staggered currents that fluctuate but do not order \cite{wen}. Varma has
proposed currents which alternate in the unit cell but do not break
translational symmetry \cite{varma}. Emery and Kivelson
\cite{emery,kivelson} and Caprara {\it et al.} \cite{caprara} have
proposed states with broken symmetries of different kinds. Our strategy
for constructing a theory and confronting experiments differs from most
others in deemphasizing modeling of the ``strange metal'' behavior and
focusing on order, low-temperature phenomenology, and material
imperfection - all issues with sharp experimental dichotomies amenable to
falsification. DDW order can be detected if it is present.  If it is not
present, the proposal is disproved.


\begin{figure}
\epsfbox{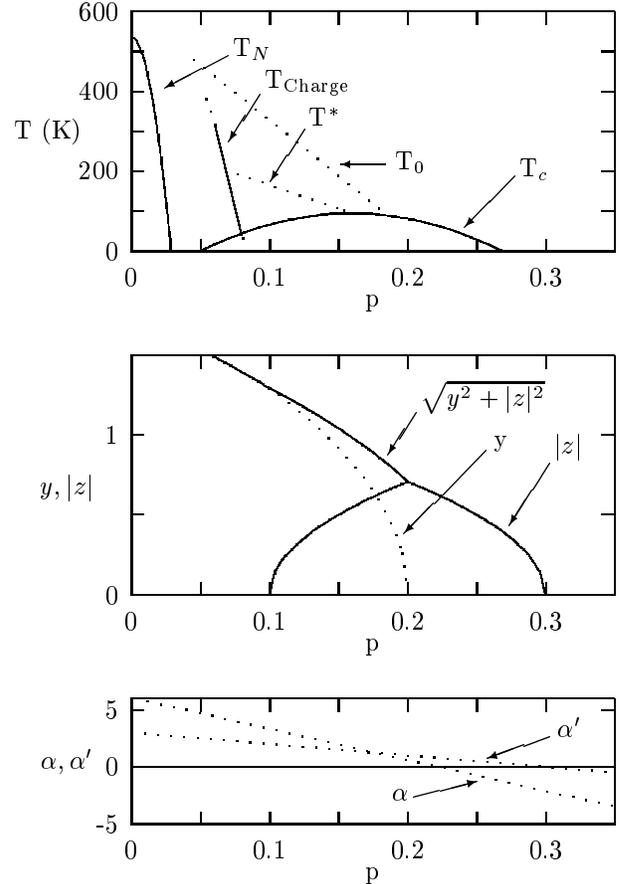}
\caption{Top: Phase diagram constructed from experiments on
         YBa$_2$Cu$_3$O$_{7-\delta}$. $T_N$ is the N\'{e}el transition,
         $T_c$ is the superconducting transition, $T^*$ is the pseudogap
         crossover [27], T$_0$ is the location of the maximum in the
         uniform susceptibilty, and $T_{\rm Charge}$ is a charge-ordering
         line recently reported by Mook {\it et al.} [39]. Middle:
         Values of $|z|$ (solid), $y$ (dots) and $(y^2+ |z|^2)^{1/2}$
         (solid) minimizing the free energy of Eq. (1) with $\lambda = 1$,
         $\gamma = -0.8$, and the linear functions $\alpha$ and $\alpha'$
         shown on the bottom.  The parameter $p$ is hole doping.}
\end{figure}

\pagebreak

\begin{figure}
\epsfbox{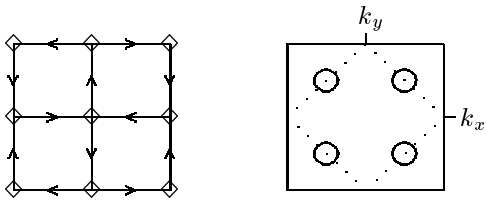}
\caption{Left: Arrangement of bond currents in the DDW state.  Right:
        Brillouin zone of the Cu-O plane.  The dots show the half-filling
        Fermi surface, as well as the Brillouin zone boundary after the
        DDW state has formed.  The circles are the Fermi surface of the
        DDW state at finite doping.}
\end{figure}

\enlargethispage*{1000pt}

\section{Competing Order}

Order-parameter competition has always been a natural candidate for
explaining why the superconducting transition temperature $T_c$ first
grows and then retreats as doping is reduced.  Let us consider the generic
zero-temperature Ginzburg-Landau free energy

\begin{equation}
F = \lambda (y^2 + |z|^2)^2 + \gamma y^2 |z|^2
 - \alpha y^2 - \alpha' |z|^2 \; \; \; ,
\end{equation}

\noindent
describing the development of order parameters $y$ and $z$ in the case
that low-order mixing is forbidden by symmetry. In Fig.  1b we plot the
values of $y$ and $z$ that minimize $F$ for the case of $\lambda = 1$ and
$\gamma =-0.8$ as a function of the abstract tuning parameter $p$.  The
variables $\alpha$ and $\alpha'$ are the simple linear functions of $p$
shown in Fig. 1c. One sees that $z$ develops at $p = 0.3$, $y$ develops at
$p = 0.2$, and that $ 0.1 < p < 0.2$ is a coexistence region in which the
growth of $y$ suppresses and eventually eliminates $z$. Thus if we imagine
$z$ to be the magnitude of the order parameter for $d$-wave
superconductivity and $p$ to be doping, then we can understand the onset,
growth, saturation, and eventual destruction of superconductivity with
reduced doping as an effect of a monotonically strengthening $d$-wave
pairing interaction, as opposed to one that first strengthens and then
weakens. The underdoped side of the superconducting dome is then
fundamentally different from the overdoped side in that the superfluid
density is suppressed there by the development of a second order parameter
$y$.

\section{D-Density Wave}

Let us now consider the order parameter

\begin{equation}
y = i \sum_{{\bf k} , s} f({\bf k})
\;  < \! c^{\dagger}_{ {\bf k+Q} , s} c_{ {\bf k} , s } \! >
\; \; \; ,
\label{ydef}
\end{equation}

\noindent
where $f({\bf k}) = \cos(k_x) - \cos(k_y)$.  If $f({\bf k})$ were replaced
by a function with $s$-wave symmetry, $y$ would simply be the order
parameter of a charge-density wave (CDW) - hence, we call this state a
$d_{{x^2}-{y^2}}$ density wave state (DDW) \cite{nayak}. For the
particular case of ${\bf Q}=(\pi,\pi)$, which we think\linebreak\pagebreak

\noindent 
most relevant to the cuprates, the equivalence of ${\bf Q}$ and $-{\bf Q}$
enforced by the underlying band structure requires the sum to be
imaginary.  Thus this state necessarily breaks parity and time-reversal
symmetry (i.e. exhibits magnetism), as well as translation by one lattice
spacing and rotation by $\pi/2$.  It is, however, symmetric under the
combination of any two of these operations.  The order parameter is
equivalent to the array of bond currents illustrated in Fig. 2.

The excitation spectrum of the DDW at very low energies is generic and
consists of conventional fermionic particles and holes in a band structure
like that of the $d$-wave superconductor with which it competes. 
Introducing a mean-field ansatz \cite{nayak} ({\it cf.} Eq. \ref{ydef}) we
obtain the 1-body Hamiltonian

\begin{equation}
{\cal H} = {\sum_{{\bf k},\sigma}} \epsilon({\bf k})
{c^\dagger_{{\bf k}\sigma}}{c_{{\bf k}\sigma}}
+ \Delta({\bf k})
{c^\dagger_{{\bf k}\sigma}}{c_{{\bf k+Q}\sigma}}
\label{mf_Ham}
\end{equation}

\noindent
where $\epsilon_{\bf k} = -2t[ \cos(k_x) + \cos(k_y)]$ and $\Delta_{\bf k}
= y V [ \cos(k_x ) - \cos( k_y )]$, and $V$ is a coupling constant in the
microscopic Hamiltonian. Microscopic Hamiltonians with short-range
repulsion and superexchange are favorable for such order
\cite{affleck,kotliar} but are even more favorable for an
antiferromagnetic state.  However, correlated hopping terms tend to tip
the balance in favor of DDW order \cite{nayak}. Since the ordering occurs
at ${\bf Q}=(\pi,\pi)$, it is most favorable at half-filling or low
doping. The corresponding band structure is

\begin{equation}
E_{\bf k} = \pm \sqrt{ \epsilon_{\bf k}^2 +
| \Delta_{\bf k}^{\rm DDW}|^2}
\;  \; \; .
\end{equation}

\noindent
At half-filling there are gapless quasiparticles only at the nodal points
${\bf k} = (\pm \pi/2, \pm \pi/2)$.  At finite doping, Fermi pockets are
opened, as shown in Fig. 2.  While the DDW state is semimetallic at
half-filling, it is a conventional metal (with 2-d localization prevented
by interlayer tunneling) with a disconnected Fermi surface at dopings
other than half-filling. It is possible for the DDW to discommensurate,
thereby opening a full gap, as occurs with a traditional spin density
wave, but this is not automatic because the remaining Fermi surface is not
nested. Some related density-wave states are discussed in the Appendix. 

The excitation spectrum at high energies is not generic. There is no
reason for the quasiparticle at $(\pi , 0)$ to have integrity,
particularly if the system is near the continuous quantum phase transition
at $p = 0.2$ in Fig. 1.  This is a Fermi-surface reconnection, at which
the Hall conductance jumps, a van Hove singularity develops at $(\pi ,
0)$, and quasiparticles scatter violently even at low energy scales
\cite{schmalian,dzyaloshinskii,morr}.

\section{D-wave Superconductivity}

The Heisenberg exchange nominally responsible for DDW order also tends to
favor $d$-wave superconductivity. This is the underlying reason the band
structures of the two are so similar, and why the competition of these two
kinds of order is natural. If we allow the superconducting bond
expectation value $< \!  c_{j \uparrow} c_{k \downarrow} \! > = \pm z$ to
develop, where the sign is positive on $x$ bonds and negative on $y$
bonds, the Hartree-Fock Hamiltonian becomes

\begin{equation}
{\cal H}_{\rm HF}' = {\cal H}_{\rm HF} + J \sum_{< j k >} \pm
( z \;  c_{k \downarrow}^\dagger  c_{j \uparrow}^\dagger
+ z^* \; c_{j \uparrow} c_{k \downarrow} ) \; \; \; ,
\end{equation}

\noindent
and the corresponding superconducting quasiparticle dispersion relation
becomes

\begin{equation}
E_{\bf k} = \pm \sqrt{[(\epsilon_{\bf k}^2 + | \Delta_{\bf k}^{\rm
DDW}|^2
)^{1/2} \pm \mu ]^2 + | \Delta_{\bf k}^{\rm DSC}|^2} \; \; \; ,
\label{dsc}
\end{equation}

\noindent
where $\Delta_{\bf k}^{\rm DSC} = zJ [ \cos(k_x ) - \cos (k_y)]$ and $\mu$
is the chemical potential.  Thus not only does this kind of interaction
stabilize both kinds of order, it allows the two order parameters to
evolve continuously into each other without collapsing the quasiparticle
gap at the zone face.  This allows us to use the ground state expectation
value of ${\cal H}$ and similar Hamiltonians as a sensible model for the
energy functional $F$, i.e. one that does not throw away important
low-energy excitations.

This calculation illustrates an important feature of the mixed state that
the superfluid density is not fixed by sum rules on the underlying Fermi
surface but is rather determined by the balance between the DDW and DSC
order parameters. This is because the superfluid is primarily a condensate
of Cooper pairs drawn from the {\it gapped} region near $(\pi , 0)$ rather
than the residual Fermi surface near $( \pi/2 , \pi/2)$. This effect is
not difficult to understand if the DDW order parameter is small, for then
the semimetallic state with Fermi points - or, away from half-filling, the
conventional metallic state with a small, disconnected Fermi surface - is
not significantly different from the parent metal with a full Fermi
surface at the energy scales relevant to superconductivity.  As the DDW
order parameter becomes large, however, we have more and more the case of
a powerful attractive force exciting electrons and holes virtually into
the insulating part of the band structure and then binding these into
superfluid.  The result is a condensate fraction that falls precipitously
as the DDW order parameter grows.  An insulating ground state (or, in this
case, nearly insulating, since there is a small disconnected Fermi
surface) that becomes a superfluid without first becoming a metal is
unusual in solids, but perhaps not in nature, for this is the central idea
behind Higgs condensation in electroweak theory.

\section{S-wave Competition}

The competition between DDW and DSC has a simple analogue in the $s$-wave
case \cite{mcmillan} that is particularly instructive because it is exact
\cite{fradkin}. The Hubbard model

\begin{equation}
{\cal H} = - t \sum_{< j k>} \sum_\sigma c_{j \sigma}^\dagger
c_{k \sigma} + U \sum_j c_{j \uparrow}^\dagger c_{j \downarrow}^\dagger
c_{j \downarrow} c_{j \uparrow}
\end{equation}

\noindent
has the special property at half-filling that replacing the fermion
operator on a lattice site $j$, $c_{j \downarrow}$, by $(-1)^j c_{j
\downarrow}^\dagger$ (a unitary transformation at half-filling) reverses
the sign of $U$. When $U > 0$ this model has a ground state which is an
ordered antiferromagnet characterized by the expectation values

\begin{equation}
\left[ \begin{array}{c}
< \! S_j^x \! > \\
< \! S_j^y \! > \\
< \! S_j^z \! >
\end{array} \right]
= \frac{1}{2}
\left[ \begin{array}{c}
< \! c_{j \uparrow}^\dagger c_{j \downarrow} +
c_{j \downarrow}^\dagger c_{j \uparrow} \! > \\
i < \! c_{j \uparrow}^\dagger c_{j \downarrow} -
c_{j \downarrow}^\dagger c_{j \uparrow} \! > \\
< \! c_{ j \uparrow}^\dagger c_{j \uparrow} -
c_{j \downarrow}^\dagger c_{j \downarrow} \! >
\end{array} \right] \; \; \; .
\end{equation}

\noindent
When $U <0$ the ground state is thus a degenerate mixture of $s$-wave
superconductivity and checkerboard charge order characterized by the
expectation values

\begin{equation}
\left[ \begin{array}{c}
< \! {\rm Re} (\Delta_j ) \! > \\
< \! {\rm Im} (\Delta_j ) \! > \\
< \! n_j \! >
\end{array} \right]
=
\left[ \begin{array}{c}
< \! c_{j \uparrow}^\dagger c_{j \downarrow}^\dagger +
c_{j \downarrow} c_{j \uparrow} \! > \\
i < \! c_{j \uparrow}^\dagger c_{j \downarrow}^\dagger -
c_{j \downarrow} c_{j \uparrow} \! > \\
(-1)^j < \! c_{ j \uparrow}^\dagger c_{j \uparrow} +
c_{j \downarrow}^\dagger c_{j \downarrow} \! >
\end{array} \right]
\end{equation}

\noindent
Both kinds of order occur simultaneously, are equivalent energetically,
and may be rotated into each other by analogy with spin rotation of an
antiferromagnet.  More precisely, this system lies at a quantum phase
transition between the two kinds of order and can be made to acquire one,
the other, or a mixture of the two by means of an arbitrarily small
perturbation, exactly the way the parameters $\gamma$ and $\alpha -
\alpha'$ in Eq. (1) break the rotational invariance of $F$.

The Hartree-Fock solution, which is only approximate, also has this
symmetry.  Allowing the expectation values $y = (-1)^j < \! c_{j
\uparrow}^\dagger c_{j \uparrow} - c_{j \downarrow}^\dagger c_{j
\downarrow} \! > / 2$ and $z = < \! c_{j \uparrow} c_{j \downarrow} \!>$
we obtain for the Hartree-Fock Hamiltonian

\begin{displaymath}
{\cal H}_{\rm HF} = - t \sum_{< j k>} \sum_{ j \sigma}
c_{k \sigma}^\dagger c_{k \sigma} + U \sum_j
\biggl[ (-1)^j y (c_{j \uparrow}^\dagger c_{j \uparrow}
\end{displaymath}

\begin{equation}
- c_{j \downarrow}^\dagger c_{j \downarrow}) + (z \; c_{j
\downarrow}^\dagger
c_{j \uparrow}^\dagger + z^* \; c_{j \uparrow} c_{j \downarrow} )
\biggr] \; \; \; ,
\end{equation}

\noindent
and for the corresponding quasiparticle dispersion relation

\begin{equation}
E_{\bf k} = \pm \sqrt{[(\epsilon_{\bf k}^2 + | \Delta^{\rm CDW} |^2
)^{1/2} \pm \mu ]^2 +| \Delta^{\rm SSC}|^2} \; \; \; ,
\end{equation}

\noindent
where $\Delta^{\rm CDW} = y U$ and $\Delta^{\rm SSC} = z U$, are the
charge density wave and $s$-wave superconducting gaps, respectively. This
is exactly the same as Eq. (\ref{dsc}) with $s$-wave quantities
substituted for $d$-wave ones. Thus, as in the $d$-wave case, the
superconducting order parameter may, at half-filling ($\mu = 0$) be
rotated continuously from pure superconductivity to pure checkerboard
charge order without closing the quasiparticle gap.  In this case,
however, the rotation also leaves the ground state energy invariant, and
is an exact symmetry \cite{zhang,fradkin}.

This calculation illustrates the important feature of charge order that
it competes easily and naturally with $s$-wave superconductivity but {\it
not} with $d$-wave.  This is because it is an $s$-wave condensate, per Eq.
(\ref{ydef}).

\section{Pseudogap}

A large number of experimental properties of the cuprates are consistent
with the presence of DDW order in underdoped samples.

\subsection{Gap Evolution}

The $d$-wave superconducting gap in the electron spectral function evolves
continuously with underdoping into the $d$-wave-like pseudogap without
collapsing. In the top of Fig. 3 we reproduce point-contact tunneling
measurements on underdoped YBCO of Renner {\it et al.} \cite{renner}
showing the excessive size of the tunneling gap and its persistence above
the superconducting $T_c$, both of which are characteristic of underdoped
cuprates. Identification of this feature with the $d$-wave gap follows
from its evolution out of the simpler BCS-like gap found in overdoped
materials and its rough compatibility with the magnitude of $T_c$.  
However, its persistence above $T_c$ is not consistent with a traditional
BCS gap, for this should disappear at $T_c$, as occurs in overdoped
samples, on quite general grounds.  That this gap has the correct angular
dependence is shown in the middle of Fig. 3, where we reproduce
angle-resolved photoemission spectra from underdoped
Bi$_2$Sr$_2$CaCu$_2$O$_{8+\delta}$ at two different points in the
Brillouin zone reported by Norman {\it et al.} \cite{norman}. The upper
trace, taken from near the zone face at $(\pi , 0)$, shows a large gap
that persists well above $T_c$, whereas the lower trace, taken from near
the node at $(\pi / 2, \pi / 2)$, shows a smaller gap that is destroyed at
$T_c$. This angular dependence is also seen in the bottom of Fig. 3, where
we reproduce the retreat of the photoemission ``leading edge'' as a
function of position on the weak-coupling Fermi surface reported by Harris
{\it et al.} \cite{Harris}. The $d$-wave-like character of the gap is
clear, as is its persistence at the zone face above $T_c$ for even
slightly underdoped samples. Thus it appears that the pseudogap and the
superconducting gap have identical functional forms and evolve
continuously into each other as the doping is reduced, just as expected
from order-parameter rotation.

\enlargethispage*{1000pt}

Energetic competition as the cause of this rotation is suggested by the
similarity between the superconducting and pseudogap energy scales.  It
may be seen in the bottom of Fig. 3 that the maximum ``leading-edge''
gapis 30 meV while the retreat caused by heating above $T_c$ is
\linebreak\pagebreak

\begin{figure}
\epsfbox{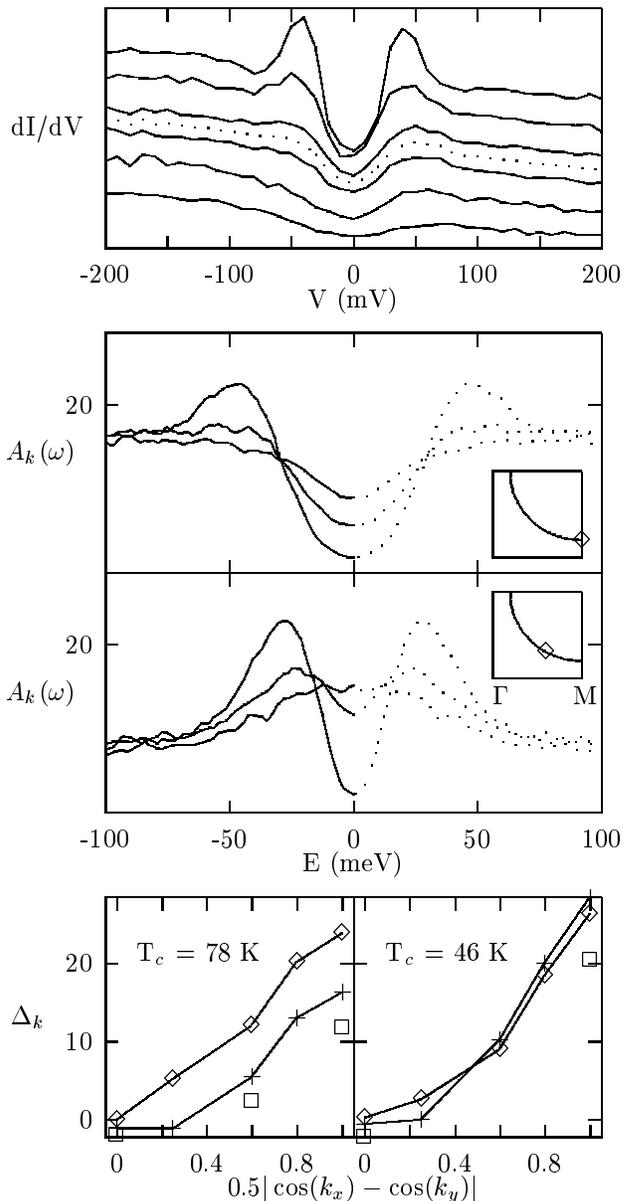}
\caption{Top: Point-contact tunneling spectrum of underdoped Bi2212 by
         Renner {\it et al.} [24] at temperatures (top to bottom) 4.2 K,
         63 K, 81 K, 89 K, 109 K, and 151 K.  The dotted line shows the
         spectrum at $T_c$ = 85 K. Middle:  Angle-resolved photoemission
         from Norman {\it et al.} [25] on underdoped Bi2212 with $T_c$ =
         75 K at the two points in the  Brillouin zone shown in the
         insets. The data were symmetrized to remove the Fermi function.
         The temperatures are from top to bottom: 65 K, 85 K, 110 K.
         Bottom: Leading-edge gap from Harris {\it et al.} [26]. Left
         panel: $T_c=78$ K Dy-BSCCO, at temperatures 13 K (diamonds), 100
         K (crosses), 150 K (squares). Right panel: $T_c=$ 46 K Dy-BSCCO;
         the symbols are the same as in the left panel.}
\end{figure}

\enlargethispage*{1000pt}

\noindent
between 5 meV
and 10 meV, depending on doping.  The pseudogap scale k$_B$T$^* \simeq$ 30
meV is also identified in a number of other measurements \cite{timusk},
notably neutron\linebreak\pagebreak

\begin{figure}
\epsfbox{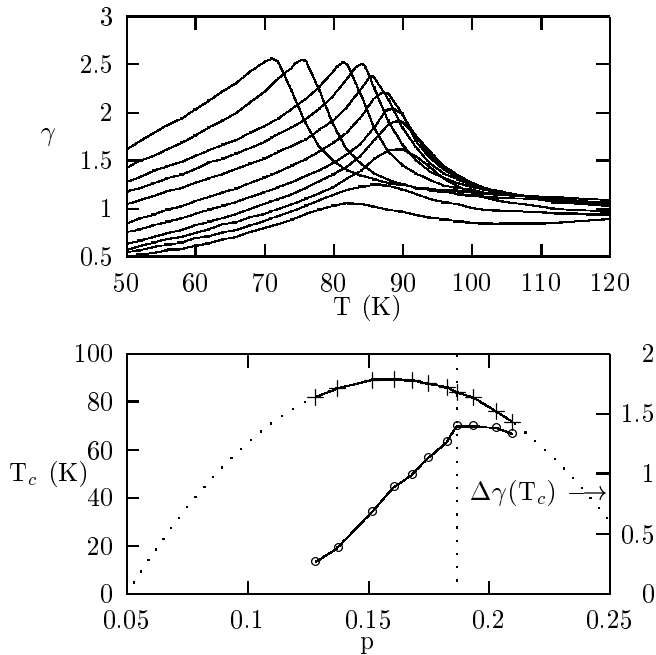}
\caption{Top: Specific heat $\gamma$ in mJ/mole K$^2$ for various doping
        levels of Bi$_{2.15}$Sr$_{1.85}$CaCu$_2$O$_{8+\delta}$ of Tallon
        and Loram [31]. The $y$-intercepts of the curves increase with
        doping so that the lower curves correspond to underdoped crystals
        while the upper curves correspond to overdoped crystals;
        the fourth highest curve corresponds to optimal doping. Bottom:
        Specific heat jump [$\gamma$($T_c$) - $\gamma$(120 K) ] of above
        samples versus doping.  The doping level is determined from $T_c$
        and the semiempirical relation relation between this and doping 
        $p$ in holes per Cu site shown in plusses.}
\end{figure}

\noindent
scattering \cite{tstar}, NMR \cite{nmr}, electronic Raman
scattering \cite{raman}, and optical reflectivity \cite{optics}.

\subsection{Superfluid Density}

\enlargethispage*{1000pt}

Rapid collapse of superfluid density below optimal doping is seen in many
experiments \cite{loram}.  The zero-temperature penetration depth, for
example, grows rapidly in the pseudogap regime and correlates with the
suppression of $T_c$ with underdoping, yet saturates at overdoping in a
way reminiscent of a traditional BCS superconductor \cite{uemura}. In Fig. 
4 we reproduce the heat capacity measurements on
Bi$_{2.15}$Sr$_{1.85}$CaCu$_2$O$_{8+\delta}$ recently reported by Tallon
and Loram \cite{loram}. Above a\linebreak hole concentration of about $p =
0.19$ per Cu the specific heat jump at the superconducting transition
varies weakly with p, as one would expect if the material were an ordinary
metal undergoing a transition to BCS superconductivity.  At $p = 0.19$,
however, there is an abrupt
transition and a rapid decrease of this height
with underdoping, as though all or part of the Fermi surface were being
destroyed by the removal of holes. As a result of this, there are fewer
low-energy excitations remaining to\linebreak\pagebreak

\begin{figure}
\epsfbox{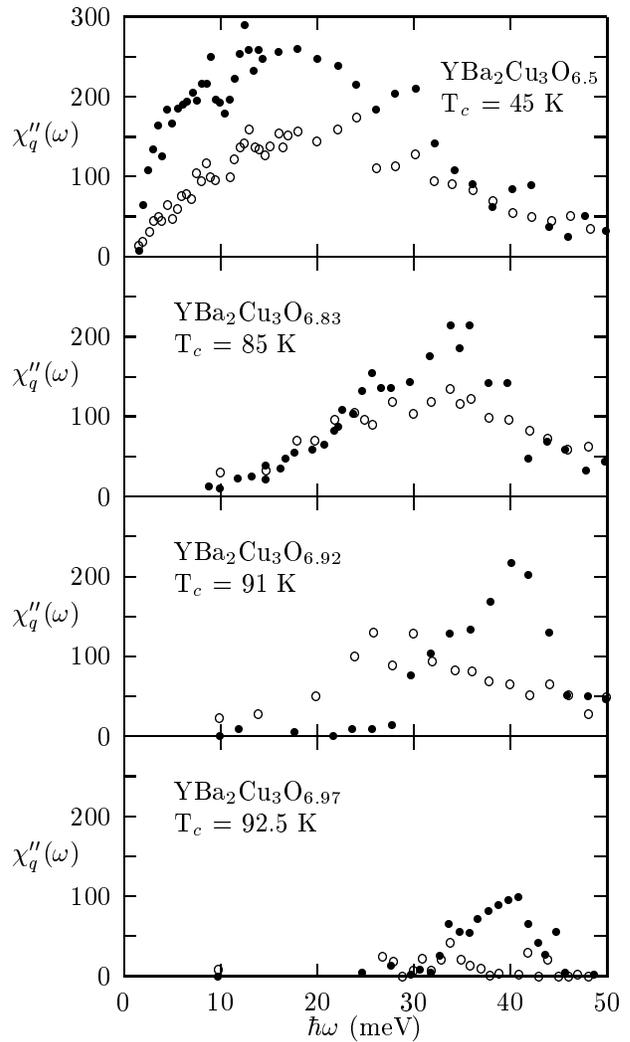}
\caption{Inelastic neutron scattering at $k = (\pi,\pi,\pi)$ reported by
        Bourges [33] for YBCO at various levels of underdoping.  The
        closed and open circles correspond to T = 5K and T = 100K,
        respectively.  Higher-resolution experiments [30] have now shown
        that the shoulder on the low-energy side below T$_c$ is actually
        split incommensurate peaks.}
\end{figure}

\enlargethispage*{1000pt}

\noindent
be affected by the superconducting transition. Hence, the specific heat
jump at the transition, $\Delta\gamma$, is reduced. All of this behavior
is compatible with Fig. 1 if the phase transition is at $p = 0.2$, where
the order parameter $y$ begins to develop, is associated with the onset of
DDW order and the consequent continuous opening of a gap at $(\pi , 0)$ in
the quasiparticle spectrum.

\subsection{Spin Susceptibility}

There is evidence that spin ordering - and thus presumably stripe ordering
- has not taken place at optimal doping in YBCO, but only occurs at much
lower doping levels. In Fig. 5 we reproduce the inelastic neutron
measurements\cite{bourges} for optimally-doped and
underdoped\linebreak\pagebreak

\begin{figure}
\epsfbox{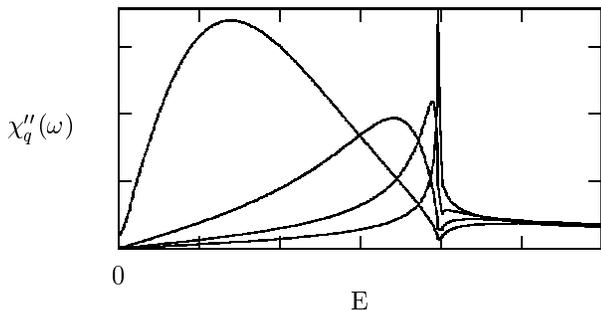}
\caption{${\rm Im} \; \chi_{\bf q} (\omega)$ as defined by Eq. 13 for the
         case of $\mu = 0$, $yJ/(2t+xJ) = 0.1$ and $U/U_c = 0.0, 0.3,
         0.6, 0.9$, where $U_c$ is the critical value of $U$.}
\end{figure}

\noindent
YBCO at a momentum transfer of $( \pi , \pi, \pi)$.  These experiments
show that the 41 meV resonance, which disappears above $T_c$ and is
presumably associated with the superconductivity, evolves continuously
with underdoping into the magnetic fluctuation spectrum of the ordered
antiferromagnet.  Thus, we interpret the piling up of low-frequency
spectral weight in the experiment at low doping as signaling the approach
of magnetic order, and conversely of showing that magnetic order is
neither present nor imminent at the onset of DDW order.  The
spin-fluctuation spectrum in the superconducting region remains fully
gapped and has no low-energy structure of any kind.  The resonance
continues to be destroyed by elevated temperature, but the requisite
temperature {\it grows} with underdoping even as $T_c$ is evolving to
zero.  In this way an excitation manifestly associated with the
superconductivity at optimal doping transforms into an excitation
irrelevant to superconductivity\cite{Chakravarty1}. 

This effect is simply understood as a triplet {\it exciton}
\cite{chubukov} that vanishes at elevated temperature because the
quasiparticle gap required for it to be well-defined vanishes.  This is
quantified in Fig. 6, where we plot the imaginary part of

\begin{equation}
\chi_{\bf q} (\omega) = \frac{\chi_{\bf q}^0 (\omega)}{1 + U \;
\chi_{\bf q}^0 (\omega)}
\; \;\ \; ,
\end{equation}

\noindent
where

\begin{displaymath}
\chi_{\bf q}^0 (\omega) = \frac{1}{2\pi^2} \int_{-\pi}^\pi
\int_{-\pi}^\pi
dk_x dk_y  \frac{E_{\bf k} + E_{\bf k+q}}
{(\omega + i \eta)^2 - (E_k + E_{\bf k+q})^2}
\end{displaymath}

\begin{equation}
\times \left( 1-\frac{\varepsilon_{\bf k} \varepsilon_{\bf k+q}
+\Delta^{DDW}_{\bf k} \Delta^{DDW}_{\bf k+q}
+\Delta^{DSC}_{\bf k} \Delta^{DSC}_{\bf k+q}}
{E_{\bf k} E_{\bf k+q}}  \right)
\end{equation}

\noindent
at ${\bf q} = (\pi,\pi)$ for various values of $U$.  This is a
crude ladder sum in which $\chi_{\bf q}^0 (\omega)$ represents the
susceptibility of the ideal BCS superconductor characterized by $E_{\bf
k}$, $\Delta_{\bf k}^{\rm DDW}$, and $\Delta_{\bf k}^{\rm DSC}$ per Eq.
(\ref{dsc}), while $U$ represents a coulomb interaction added to push this
system toward spin antiferromagnetism.  One sees that as $U$ is increased
the sharp resonance in the spectrum decreases in energy and broadens, just
as occurs with decreased doping in Fig. 5.  This width is due to efficient
decay of the exciton into nodal quasiparticle pairs. At a slightly higher
value of $U$ the continuum evolves into a divergence at $\omega = 0$
associated with onset of spin order.  Note that the DDW and DSC order
parameters in this calculation are effectively interchangeable. Since
$\Delta_{\bf k}^{\rm DDW} = - \Delta_{\bf k+q}^{\rm DDW}$ and $\Delta_{\bf
k}^{\rm DSC} = - \Delta_{\bf k+q}^{\rm DSC}$ for ${\bf q} = (\pi,\pi)$,
the coherence factor is unity and unchanged close to the Fermi energy
whether or not both gaps, or only one of them, are present.  For an
$s$-wave gap the corresponding coherence factor would have been zero.

\subsection{High-Field Transport}

Stripes and antiferromagnetic order are naturally associated with the
insulating behavior of the cuprates seen near half-filling \cite{zaanen}. 
In a conventional doped band insulator, insulation is caused by
impurities, which trap carriers and prevent them from moving.  The system
becomes a metal when it is doped sufficiently that the impurity orbitals
touch.  One of the most significant characteristics of the cuprates is
that they continue to insulate to phenomenally high dopings, typically 5\%
or 1 hole for every 20 Cu atoms.  It is very difficult to understand how
an insulator with an energy gap less than that of the common semiconductor
GaAs should still insulate at these high dopings through impurity trapping
solely. But development of antiferromagnetic order with antiphase domain
walls, which then trap carriers and pin, is easy to understand, physically
sensible, and supported experimentally by the simultaneous occurrence in
these materials of discommensurated magnetic Bragg peaks and X-ray
satellites at exactly half their momentum displacements \cite{tranquada}. 
Thus our view is that charge ordering (which would have an order parameter
of the form (\ref{ydef}), but with an $f({\bf k})$ which has $s$-wave
symmetry and, in all likelihood, incommensurate ${\bf Q}$)  impedes
conduction, rather than facilitating it \cite{emery}, and moreover is
characteristic of the insulating state. 

\enlargethispage*{1000pt}

The issue of coexistence of superconductivity with stripes and
antiferromagnetism, and potential causative relations among them, is still
highly controversial and a matter of experimental study \cite{shirane}.
There is, however, increasing evidence that the coexistence found in
La$_{2-x}$Sr$_x$CuO$_4$:Nd \cite{tranquada} is anomalous and that the
cuprates with the highest values of $T_c$ have charge ordering only at the
low-doping edge of the superconducting dome.  The recent neutron
scattering from YBa$_2$CuO$_3$O$_{7-x}$ reported by Mook {\it et al.}
\cite{mook} find the charge-ordering line shown in Fig. 1 and no static
antiferromagnetism anywhere in the superconducting region.  This is
consistent with the the high-field transport experiment on
Bi$_2$Sr$_{2-x}$La$_x$CuO$_{6+\delta}$ recently reported by
Ono\linebreak\pagebreak

\begin{figure}
\epsfbox{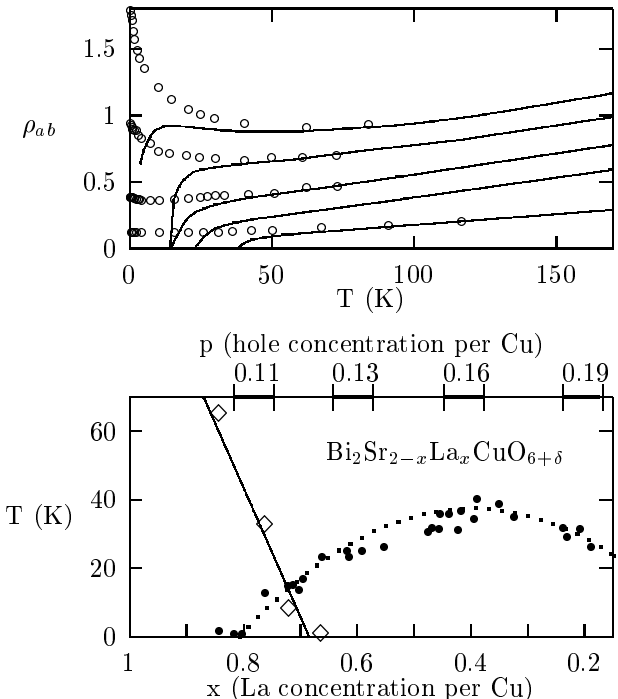}
\caption{High-field transport experiment of Ono {\it et al.} [40] on
         Bi$_2$Sr$_{2-x}$La$_x$CuO$_{6+\delta}$.  Top: Resistivity
         in m$\Omega$-cm versus temperature for samples with La
         concentrations (top to bottom) x= 0.84, 0.76, 0.73, and 0.39
         with (circles) and without (solid lines) an applied magnetic
         field of 60 T.  Bottom: The temperature of the minimum value of
         $\rho$ (diamonds) plotted versus hole concentration.  The
         superconducting T$_c$ (dots) is plotted for reference.}
\end{figure}

\noindent 
{\it et al.} \cite{ono} reproduced in Fig. 7, which finds a
metal-insulator transition at essentially the same doping as the charge
ordering line of \cite{mook} when the superconductivity is crushed by a
large magnetic field.  The phenomenology of this transition is
qualitatively similar to that observed previously in
La$_{2-x}$Sr$_x$CuO$_4$ \cite{boebinger} except that it occurs near the
edge of the dome rather than near optimal doping.  This is important, for
Castellani, Di Castro, and Grilli \cite{castellani} were led by this
observation to propose that the strange-metal behavior of the cuprates
might be quantum criticality associated with the charge-ordering
transition. These more recent experiments suggest that it is instead
quantum criticality associated with the development of DDW order. LSCO is
unique among the high-T$_c$ cuprates in having a low transition
temperature, a strong tendency to stripe-order near 1/8 doping, and an
extreme sensitivity to Nd doping \cite{tranquada}, all of which suggest
mechanical weakness of the crystal structure. 

The large-field experiment also reveals another important aspect of the
cuprates, namely the lack of evidence for strange-metal behavior in the
\enlargethispage*{1000pt}
zero-temperature normal state. It may be seen in the top of Fig. 7 that
the resistivities on the metallic side of the transition become constant
at low temperatures and that they evolve\linebreak\pagebreak

\begin{figure}
\epsfbox{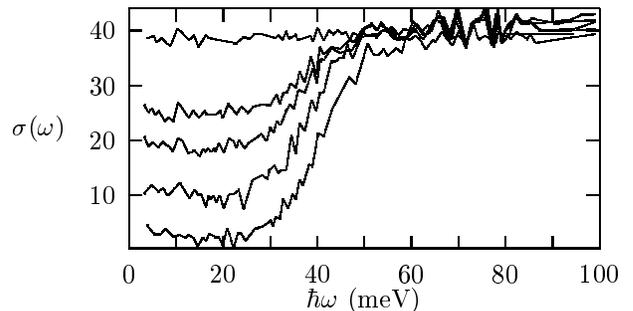}
\caption{$c$-axis optical conductivity of YBa$_2$Cu$_3$O$_{6.7}$ in 
        $\Omega^{-1} {\rm cm}^{-1}$ reported by Homes {\it et al.} [43]
        for temperatures (top to bottom) T = 250 K, 150 K, 110 K, 70K,
        and 10 K . Large phonon contributions have been subtracted
        out.}
\end{figure}

\noindent
continuously across the transition into the linear-$T$ resistivity
characteristic of the high-temperature normal state of the cuprates.  The
resistivity at the transition is also about 200 $\mu \Omega$-cm, a typical
saturation resistivity in strong-scattering metals.  Both of these
properties are consistent with the zero-temperature normal state being a
conventional metal.  They do not prove this, but they make the argument
for a non-Fermi-liquid phase more difficult, as linear-$T$ resistivity is
one of its key signatures. 

Thus on the basis of these experiments we predict that in large magnetic
fields there should be a {\it second} zero-temperature phase transition
near $ p = 0.19$ associated with the onset of DDW order.  At this
transition the system should remain a conventional metal but violently
change the topology of its Fermi surface.  This transition should be
plainly visible in all transport measurements and should be characterized
by powerful critical scattering.\cite{schmalian,dzyaloshinskii,morr}

\subsection{$c$-Axis Conductivity}

DDW formation provides a simple explanation for the perplexing
semiconducting $c$-axis resistivity in many cuprates.  In Fig. 8 we
reproduce the optical conductivity measurements on YBa$_2$Cu$_3$O$_{6.7}$
of Homes {\it et al.} \cite{homes} showing the steady reduction of the
oscillator strength below 40 meV beginning at a temperature far above the
superconducting T$_c$.  That this reduction does not conserve the f-sum
rule locally - which any mean-field theory, including that of the DDW,
does - is interesting but not necessarily significant, as a mean-field
description is obligated to be quantitative only at arbitrarily small
energy scales. Band structure studies \cite{ollie} of these materials have
shown that the $c$-axis tunneling matrix element is largest at $(\pi,0)$
and symmetry-related points - precisely at the points where the DDW gap is
large.  (The functional form is roughly $t_\perp \sim (\cos k_x - \cos
k_y)^2$ \cite{ollie}.) Thus the opening of the DDW gap suppresses the
$c$-axis transport because the remaining Fermi surface does not conduct
efficiently in the c-direction due to small tunneling matrix element.  The
above matrix element holds for simple tetragonal materials (Hg1201,
Tl1201, etc.). For body centered tetragonal materials (LSCO, Tl2201,
Bi2212, etc.), the maxima of $t_\perp$ are shifted towards the zone
center, and the effect of opening the DDW gap at $(\pi , 0)$ and
symmetry-related points is weaker.

\section{Orbital Magnetism}

The distinguishing characteristic of DDW order is the magnetic field it
makes. Since the possibility of spontaneous breaking of time-reversal and
parity in the cuprates was first proposed in the late 1980s there have
been a number of attempts to detect such fields, most of which have
reported null results \cite{kiefl}.  However there has always been
confusion about the size of the effects one would expect, and there have
always been mysterious magnetic signals in the cuprates, including a
recent report of spin antiferromagnetism coexisting with superconductivity
in a sample of superoxygenated La$_2$CuO$_{4+y}$ with y=0.12 and T$_c$ =
42 K \cite{shirane}.  This fundamentally conflicts with a previous report
of no magnetism in La$_{1.85}$Sr$_{0.15}$CuO$_4$ \cite{kiefl}.  We feel
that the magentic experiments are so contradictory that they can at
present neither rule out nor confirm the presence of DDW order.

We estimate the magnetic field at the center of a plaquette associated
with DDW order to be between 1 and 30 gauss \cite{hsu}.  The bond currents
of Fig. 2 are roughly $e\Delta^{\rm DDW}/\hbar$, where $\Delta^{\rm DDW}$
is the maximum DDW gap.  If we take this to be 30 meV, we find bond
currents of about $7 \mu A$.  The large uncertainty in the corresponding
field strength is due mainly to uncertainty in the current path.  One can
reasonably consider models ranging from Cu sites connected by 1 \AA $\;$
``wires'' to split current carried between adjacent O atoms.

Let us now briefly review the current experimental situation relevant to
direct detection of DDW magnetism.

\subsection{Neutron Scattering}

DDW order is, in principle, visible in magnetic neutron scattering.
Unfortunately the signals are quite small compared with those from ordered
spins and easily overwhelmed by them. The ratio of the staggered magnetic 
field associated with DDW fields to that nominally produced by an ordered
array of spins is

\begin{equation}
\frac{B_{\rm DDW}}{B_{\rm AFM}} = \left(\frac{ e
\Delta^{\rm DDW}}{\hbar c r}\right)  \left(\frac{m c r^3}{e \hbar}\right)
= \frac{m r^2}{\hbar^2} \Delta^{\rm DDW} \; \; \; ,
\end{equation}

\noindent
or about 0.06, with $r = 4$ \AA $\;$ taken for the bond length.
Effective magnetic moments of this size are just barely detectable
in the cuprates \cite{shirane,birgeneau}.

It is also unfortunate that the doping levels at which DDW order should be
well developed lie close to the spin-glass regime \cite{chou} where the
system crosses over between N\'{e}el and superconducting order.  The spin
glass is characterized by slightly incommensurated short-range
antiferromagnetism with strongly suppressed scattering intensities along
one orthorhombic axis - behavior consistent with unpaired Cu spins
pointing in the plane \cite{birgeneau} and {\it in}consistent with DDW
magnetism. However, numerous incursions of this magnetism into the
superconducting phase have been reported, in one case deeply
\cite{shirane}, and this has always been difficult to understand from the
point of view of traditional magnetic models.  It implicitly raises the
question of whether there might be two kinds of antiferromagnetism in the
cuprates - one, due to spins, which is incompatible with superconductivity
and one, due to DDW, which is fully compatible with it and associated with
pseudogap formation. Spin-orbit coupling would then mix these and
conceivably make them evolve into each other with increased doping.

\subsection{X-ray Scattering}

DDW order cannot be seen in X-ray scattering.  The DDW order parameter is
odd under time-reversal while atomic displacements are even, so there is
no first-order coupling between them, and Bragg scattering through
circular birefringence from the valence electrons is too weak.  For a 10
KeV X-ray of frequency $\omega$ the Bragg intensity is down by the factor
($\mu_B B^{\rm DDW}/\hbar\omega)^2 \simeq 10^{-16}$ from the Bragg
intensity of valence electrons - already small compared with the signal
from the core electrons.  The absence of an X-ray signal is a key
characteristic DDW order distinguishing it experimentally from stripes.

\subsection{Magnetic Resonance}

The static magnetic field of ideal DDW order cannot be seen directly
through NMR of Cu or O nuclei in ideal CuO planes, as these lie at centers
of symmetry where the DDW magnetic field is zero.  However, magnetic
fluctuations associated with the onset of DDW order or a glassy state of a
disorder-frustrated DDW could be seen by NMR, although it would be
difficult to distinguish from antiferromagnetic spin fluctuations for the
reasons stated above. Also, DDW order {\it can} in principle be seen in
NMR of ions out of the Cu-O planes, such as Y, Ba, La, or Sr. It has long
been established that there are unusual magnetic signals below T$_c$ in
all the cuprates, but attempts to quantify these have been plagued by the
\enlargethispage*{1000pt}
inherent model-sensitivity of NMR analysis.  Tallon and Loram
\cite{tallon} have recently argued using the ratio of $^{63}$Cu and
$^{17}$O spin-lattice relaxation rates analyzed with\linebreak\pagebreak

\begin{figure}
\epsfbox{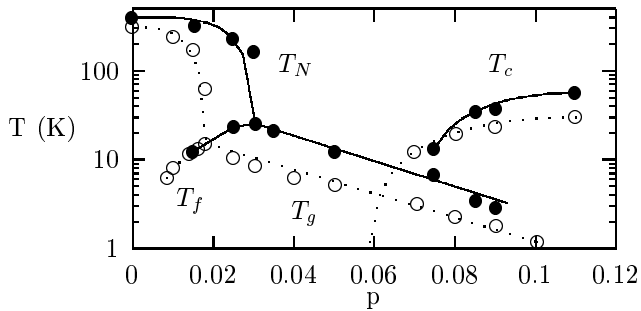}
\caption{Cuprate phase diagram inferred by  Niedermayer {\it et al.}
        [47] from $\mu$SR measurements.  The solid and open symbols refer
        to Y$_{1-x}$Ca$_x$Ba$_2$Cu$_3$O$_6$ and La$_{2-x}$Sr$_x$CuO$_4$,
        respectively.  $T_N$ is the N\'{e}el temperature, $T_c$ is the
        superconducting transition temperature, $T_f$ is a secondary
        magnetic ordering that occurs on top of spin antiferromagnetism,
        and $T_g$ is the glass transition temperature.  The dephasing
        time for the latter is approximately 0.1 $\mu$S.}
\end{figure}

\noindent 
the model of Millis, Monien, and Pines \cite{pines} that short-range
antiferromagnetic fluctuations develop in the pseudogap regime with a
functional dependence on $p$ tracking roughly the value of $y$ in Fig. 1. 
This analysis is not persuasive evidence for DDW order.

\subsection{Muon Spin Resonance}

Muon spin resonance has consistently found evidence for magnetism in the
superconducting state of the cuprates for dopings less than $p = 0.1$. In
Fig. 9 we reproduce the phase diagram of Niedermeyer {\it et al.}
\cite{budnick} showing boundaries of distinct magnetic behaviors observed
in powders of both La$_{2-x}$Sr$_x$CuO$_4$ and
Y$_{1-x}$Ca$_x$Ba$_2$Cu$_3$O$_{6.02}$. They report a ``spin freezing''
transition ($T_f$) below the N\'{e}el transition, and a spin-glass
transition ($T_g$) that cuts in to the superconducting dome. Below this
transition, and at doping levels as high as $p=0.09$, the muons depolarize
in about 0.1 $\mu$S.  In the case of LSCO, the measurements extended into
the range of the 1/8 anomaly at p = 0.12, beyond which the spin-glass
neutron signal tends to disappear \cite{birgeneau} and where no magnetism
was found in previous $\mu$SR measurements \cite{kiefl}.  However the fact
that both cuprates behave similarly, and that the spin-glass temperature
in Y$_{1-x}$Ca$_x$Ba$_2$Cu$_3$O$_6$ is substantially higher, suggests that
this behavior is characteristic of the cuprates as a class. Also, the way
the spin-glass line ends has always been confusing.

In a recent paper Panagopoulos {\it et al.} \cite{pana} have reported
anomalous long-time magnetic fluctuations at temperatures just above the
glass transition in La$_{2-x}$Sr$_x$CuO$_4$ powders.  The scale of these
is comparable to T$_c$ and has a functional dependence with doping
identical to that of the parameter $y$ in Fig. 1 - i.e. decreasing with
doping and vanishing at $p = 0.17$.  Thus they argue that the spin-glass
line actually ends here, not at 1/8.  The observation of the same effect
in a different cuprate, which seems likely in light of Fig. 9, would
suggest an intrinsic magnetic signal developing at the onset of DDW order.

\section{Disorder and Crossover}

The muon phenomenology suggests an answer to a question plaguing the idea
of competing order in the cuprates, namely why there is no evidence for a
genuine phase transition at the pseudogap temperature $T^*$, the alleged
phase boundary for onset of DDW, and also why previous searches for
magnetism at optimal doping have found sample-dependent or null results.
It is simply that the DDW order is corrupted by disorder and transformed
into the spin-glass transition at the lower temperature
$T_g$\cite{disorder}.  In very dirty samples it is lowered so much as to
be effectively destroyed.

There has been controversy over how much intrinsic disorder cuprates
possess since they were discovered.  The essence of the problem is that
the most sensitive tests of disorder - transport and the degradation of
superconductivity - are corrupted by the non-Fermi-liquid behavior of the
normal state \cite{mfl} evidenced by resistivities which exceed the
Ioffe-Regel limit of 100 $\mu\Omega$-cm at T$_c$ ({\it cf.} Fig. 7) and
increase with temperature from there. However, using criteria less
dependent on the theory of metals, the case for chronic disorder is easier
to make: All cuprates lose oxygen easily in arbitrary amounts.  All of
them have spin-glass phases at low dopings \cite{pana}.  All of them have
magnetic scattering in the superconducting regime that is
sample-dependent, difficult to reproduce, and difficult to quantify
\cite{mook}. All of them have anomalous widths in Cu and O NMR and NQR
\cite{nmrdis}.  Thus our view is that all cuprates made thus far have been
significantly disordered, even ones showing evidence to the contrary such
as narrow superconducting transition widths.  This view is supported by
new scanning tunneling microscope experiments on atomically perfect
cleaves of optimally-doped BSCCO that find inhomogeneities in the
tunneling density of states on the scale of 20 \AA \cite{seamus}.

We note that the disorder need not occur within the CuO$_2$ planes to have
a strong effect on the electronic properties. The non-superconducting
cuprates which were studied in the late 1980s differ from the
superconducting ones only in the elements that sit between planes.
Substituting Hg between the planes raises T$_c$ substantially.
Substituting Nd causes stripes \cite{tranquada}.

The DDW transition is in the same universality class as the random-bond
Ising model.  The DDW order parameter breaks translational and rotational
symmetries, and thus couples to disorder as in a model with a random
uniaxial anisotropy.  From the Imry-Ma argument \cite{disorder} as adapted
to the random anisotropy case \cite{pelcovits}, we know that this
symmetry-breaking transition will be spoiled by the random distribution of
impurities.  Thus, in the presence of disorder, time-reversal is the only
true symmetry that can be spontaneously broken by the DDW state.  The
universality class is then that of a $Z_2$ symmetry preserved by the
impurities. 

The phase diagram of the random-bond Ising model depends critically on the
disorder strength. Weak disorder is an irrelevant perturbation and can be
ignored at the finite-temperature transition to a state with broken time
reversal symmetry. Such a state has a non-vanishing expectation value for
the staggered orbital magnetization (thereby breaking the {\it
disorder-averaged} translational symmetry).  If the disorder is strong, on
the other hand, and the interlayer coupling is finite, then there can be a
finite-temperature transition in the same universality class as the
three-dimensional ($3D$) Ising spin-glass transition. Due to the weakness
of the magnetic coupling between the planes, the spin-glass transition
temperature estimated from the two-dimensional spin-glass
susceptibility\cite{singh}, $\chi_{sg}^{2D}\sim T^{-\gamma}$,
$\gamma\approx 5.3$, is small. Such a transition would not be possible if
we could neglect the coupling between the planes, as the lower critical
dimension of the Ising spin glass is known to be greater than two. 

In contrast to this, the finite temperature transition to DSC remains
sharp in the presence of disorder - although $T_c$ may be degraded. This
is because disorder does not couple linearly to the order parameter as a
random field, and because a superconducting transition in two dimensions
is possible.  The ultimate 3-dimensional transition driven by the coupling
between the layers is robust. This can be further understood by invoking
the Harris criterion \cite{disorder} assuming that the transition is in
the $3D$-XY universality class. The criterion states that weak disorder is
an irrelevant perturbation to the pure system if the specific heat
exponent is negative, which is indeed the case for the $3D$-XY model. When
the disorder is so strong that $k_F l\sim 1$, where $k_F$ is the Fermi
wave vector and $l$ is the mean free path, a superconductor-insulator
transition will take place, and the Harris criterion will no longer apply. 

The potential presence in this system of a disorder-sensitive, purely
electronic, phase transition involving a Fermi-surface reconnection raises
the disturbing possibility that many experiments in this field may be
measuring corrupted critical properties of the DDW transition rather than
the properties of new states of matter.  The notorious non-Fermi-liquid
behavior of the normal state, for example, appears to evolve at the lowest
temperatures and in a strong magnetic field into behavior of a traditional
metal.  A possible explanation of this is that the high-temperature
behavior is characteristic of a quantum critical \cite{sachdev} region
associated with a nearby critical point.

\section{Summary}

In summary we find that most of the strange behavior of the cuprate
superconductors is consistent evidence for the simultaneous occurrence of
$d$-wave superconductivity and bond antiferromagnetism.  On the basis of
this we predict that the spin-glass transition temperature observed in
muon spin resonance will climb to the pseudogap temperature $T^*$ as the
sample quality improves, that the onset of this effect with doping
coincides perfectly with the loss of superfluid density at $p = 0.19$, and
that this transition will be found to be a metal-metal transition
involving a Fermi surface reconnection, {\it not} a transition to stripe
order \cite{emery,kivelson,caprara,zaanen,tranquada}, when magnetic fields
sufficiently intense are available to crush the superconductivity at
optimal doping.

\section*{Acknowledgements}

RBL wishes to thank Z.-X. Shen, D. Pines, S.-C. Zhang, and G. Aeppli for
numerous helpful discussions.  RBL, SC, and DM wish to thank the Institute
for Complex Adaptive Matter at Los Alamos, where key ideas for this work
were conceived. RBL was supported by the National Science Foundation under
Grant No. DMR-9813899 and by NEDO. SC was supported by the National
Science Foundation under Grant No. DMR-9971138 and, in part, by funds
provided by the University of California for the conduct of discretionary
research by Los Alamos Natonal Laboratory, under the auspices of the
Department of Energy. CN was supported by the National Science Foundation
under Grant No. DMR-9983544 and the Alfred P. Sloan Foundation. DM was
supported by the Department of Energy at Los Alamos National Laboratory.

\section*{Appendix: Related Density Waves}

There are two related unconventional density wave order parameters
potentially relevant to the cuprates\cite{nayak}.  The first is that the
frustration of the singlet DDW order parameter can lead to incommensurate
ordering, in analogy with the Ferrell-Fulde-Larkin-Ovchinnikov
\cite{ferrell} state in superconductivity as nesting is destroyed. As in
the superconducting case, this will take place for sufficiently strong
frustration and at sufficiently low temperatures. Note that in this case
the order parameter is allowed to couple with lattice displacements and
can therefore be seen in X-ray scattering.  When the order parameter is
incommensurate, it will no longer have pure $d_{{x^2}-{y^2}}$ symmetry,
but will mix in $p$-wave terms. For ${\bf Q}=(\pi/a,\pi/a)+{\bf q}$ with
$|{\bf q}|$ small, the order parameter will take the form of\linebreak

\noindent 
equation (\ref{ydef}), with $f({\bf k})= \cos (k_x) -\cos (k_y)$ replaced
by \cite{nayak}

\begin{displaymath}
f({\bf k}) = (1 + \frac{i}{2} q_x a ) \left[ \cos (k_x a ) -
\cos (k_y a ) \right]
\end{displaymath}

\begin{equation}
- \frac{1}{2} q_x a \; \sin (k_x a )
- \frac{1}{2} {q_y}a \; \sin ( k_y a ) \; \; \; .
\end{equation}

The second interesting order parameter is the triplet version of the
DDW\cite{halperin,schulz,nayak}. This is defined by

\begin{equation}
\vec{y} =
\sum_k f({\bf k})\sum_{ss '} \vec{\sigma}_{ss'}
< \! c_{{\bf k+Q}, s}^\dagger c_{{\bf k}, s'} \! >
\; \; \; ,
\end{equation}

\noindent
where $\vec{\sigma}$ is a Pauli spin matrix.  If $f({\bf k})$ were chosen
to be a function of $s$-wave symmetry, this would be a conventional
spin-density wave.  The order in this case is chracterized by broken
time-reversal, translational, and rotational invariances.  The combination
of any two of time-reversal, a translation by one lattice spacing, or a
rotation by $\pi/2$ is preserved, however. In addition, spin-rotational
symmetry is also broken, which leads to gapless spin-1 excitations. The
triplet DDW corresponds to an alternating pattern of {\it spin} currents
analogous to charge currents of Fig. 2. Presently, the phenomenology of
high temperature does not seem to be consistent with the choice of the
triplet DDW as the competing order parameter.

\end{document}